\documentclass[preprint,showkeys,amsmath,amssymb]{revtex4}

\usepackage[dvipdfm]{graphicx}
\usepackage{dcolumn}
\usepackage{bm}
\usepackage{color}

\begin{document}

\title{Torque Theory of Anisotropic Superconductors with No Phenomenological Parameter in determining Vortex Core Size}


\author{Daichi Kubota$^\mathrm{A}$}
\author{Nobuhiko Hayashi$^\mathrm{B,C}$}
\author{Takekazu Ishida$^\mathrm{A,C,D}$}
\email{ishida@center.osakafu-u.ac.jp}
\affiliation{%
$^\mathrm{A}$Department of Physics and Electronics, Osaka Prefecture University, 1-1, Gakuen-cho, Naka-ku, Sakai, Osaka 599-8531, Japan
}%
\affiliation{%
$^\mathrm{B}$Nanoscience and Nanotechnology Research Center (N2RC), Osaka Prefecture University, 
1-2 Gakuen-cho, Naka-ku, Sakai, Osaka 599-8570, Japan
}%
\affiliation{%
$^\mathrm{C}$CREST(JST), Sanban-cho, Chiyoda-ku, Tokyo 102-0075, Japan}
\affiliation{%
$^\mathrm{D}$Institute for Nanofabrication Research, Osaka Prefecture University, 1-1, Gakuen-cho, Naka-ku, Sakai, Osaka 599-8531, Japan
}%

\date{\today}

\begin{abstract}

The contribution of vortex core has been taken into account properly in constructing a torque theory for multiband superconductors.
We employ the prescription of describing internal magnetic field in the vortex lattice by Hao {\it et al.} and by Yaouanc {\it et al.} to derive a torque formula as a natural extension of a preceding London theory.
In marked contrast with the preceding model, our novel formula does not contain a phenomenological parameter $\eta$, which prevents us from obtaining a {\it true} upper critical field $H_{\rm c2}$ by analyzing an experimental torque curve.
The parameter $\eta$ was originally introduced to take care of the uncertainty in determining the vortex core size $\xi_v$.
Furthermore, we reveal that the $\eta$ value is universally scaled by anisotropy $\gamma$, magnetic field $B$, and $H_{\rm c2}$ due to field dependence of $\xi_v$.
This may revitalize the single-band Kogan model in combination with a universal function $\eta(\gamma, B, H_{\rm c2})$ instead of a constant $\eta$.

\end{abstract}



\maketitle

Thanks to extensive studies on the superconductivity of a magnesium diboride and iron arsenides, the torque as the angular derivative of a free energy has been recognized as a powerful tool for investigating superconducting {anisotropy}.
For example, MgB$_2$ can be characterized by isotropic three-dimensional $\pi$ band and anisotropic two-dimensional $\sigma$ band. 
Discovery of high-$T_{\rm c}$ cuprates strongly suggested that lower dimensionality is a key prerequisite for inducing high-$T_{\rm c}$ superconductivity.
A recent study by Yuan {\it et al.} \cite{Yuan2009} is indicative of an opposite view because they found a rather isotropic superconducting anisotropy in (Ba,K)Fe$_2$As$_2$. 
As a sensing probe of anisotropy, the torque would play an important role for revealing the nature of various new superconductors.

Theoretically, the magnetic torque of an anisotropic superconductor has been investigated for a long time. 
Originating from the local-limit model of the torque, it evolved into a model, which takes care of the nonlocal effect for extending to a variety of superconductors.
In the early London model of the torque \cite{Kogan1988}, a vortex core is not considered seriously but is simply assumed to confine at the region shorter than a coherence length $\xi$.
The later Kogan model \cite{Kogan2002a} discusses not only the anisotropy of a penetration depth $\lambda$ but also that of $\xi$ independently, but it still remains to be a local model.
Yaouanc {\it et al.} \cite{Yaouanc1997} computed the Fourier components of the magnetic field in a high-$\kappa$ type-II superconductor containing an ideal vortex lattice for analyzing the local magnetic field in $\mu$SR and neutron diffraction measurements. 
Their method revealed the effect of the vortex core for determining local magnetic field.
Brandt developed an analytical method using full Ginzburg-Landau calculations for ideal vortex lattice with any spatial symmetry \cite{Brandt1997}.
The method developed by Yaouanc {\it et al.} gave a useful tool in analyzing the vortex state in NbSe$_2$ single crystals \cite{Miller2000,Sonier1997} and the effect of the multiband nature on the local field profile in $\mu$SR experiments \cite{Sonier2005}.

There is an issue to be amended in the London model because the cutoff in the reciprocal Fourier space ($G$-space) at $2\pi\xi^{-1}$ coming from the vortex core is introduced rather a priori. 
There is still room for improving the London theory since it is recognized that the core size (or its $G$-space equivalent) depends on the magnetic field. 
In this Letter, we attempt to derive the torque formula on the basis of the discrete Fourier-analysis method of the internal magnetic field developed by Yaouanc {\it et al.} \cite{Yaouanc1997}.

Before explaining our new theory, we first summarize the development of the preceding torque theories.
The torque is directly obtained by differentiating a free energy with respect to angle $\theta$ between crystalline axis and the magnetic field.
The free energy density $F$ is given by \cite{Kogan1996}
\begin{eqnarray}
F=\frac{B^2}{8\pi\Phi_0}{\sum_\mathbf{G}} h_z(\mathbf{G}) \,,
\label{Kogan-free-energy}
\end{eqnarray}
where {$h_z(\mathbf{G})$}, as in Ref.~\cite{Kogan1996}, is the component of the magnetic flux in the $G$ space, $B$ is the magnetic flux density, and $\Phi_0$ is the quantized flux. 
We take coordinates with $z-$axis parallel to magnetic field.
The magnetic flux {$h_z(\mathbf{G})$} is derived by
\begin{eqnarray}
h_z(\mathbf{G})=\frac{\Phi_0(1+\Lambda_{zz}G^2)}{(1+\Lambda_{zz}G_x^2+\Lambda_cG_y^2)(1+\Lambda_aG^2)} \,,
\label{BzG-2}
\end{eqnarray}
where $\Lambda_{ij}$ is a penetration depth tensor, $\Lambda_{ij}=\lambda^2{m_{ij}}$  ($\Lambda_a=\lambda^2{m_a}$, $ \Lambda_c=\lambda^2{m_c}$) is represented by $m_{xx}=m_a\cos^2\theta+m_c\sin^2\theta$, $m_{xz}=(m_a-m_c)\sin\theta\cos\theta$, $m_{yy}=m_a$, and $m_{zz}=m_a\sin^2\theta+m_c\cos^2\theta$ \cite{Campbell1988}.

Suppose that the relation $H_{c1} \ll H \ll H_{c2}$ is satisfied in the intermediate field so that an effective vortex spacing $L$ can be expressed as $\xi \ll L \ll \lambda$.
Using the dimensionless reciprocal lattice vectors $\mathbf{g} = L\mathbf{G}$ with $L = (\Phi_0/B)^{1/2}$, we approximate free energy density within the zeroth and first-order expansion with respect to $\mathbf{g}$.
Actually the free energy density is approximated as \cite{Campbell1988},
\begin{eqnarray}
F = \frac{B^2}{8\pi}+\frac{B^2}{8\pi}\frac{m_{zz}L^2}{\lambda^2m_a}{\sum}^\prime \frac{1}{m_{zz}g_x^2+m_cg_y^2},
\label{Kogan FE}
\end{eqnarray}
where $\sum^\prime$ sums up $\mathbf{g}$ ($\mathbf{g}\ne 0$).
A free energy density is given from Eq.~(\ref{Kogan FE}) as \cite{Campbell1988}
\begin{eqnarray}
8\pi F 
= B^2+\frac{\Phi_0}{4\pi\lambda^2}\sqrt{m_aB_x^2+m_cB_z^2}\ln(\eta H_{c2}/B)\, .
\label{Campbell}
\end{eqnarray}
where a phenomenological parameter $\eta$ ($\sim 1$) was first introduced \cite{Campbell1988} and has been anxious about the clarification of its physical meaning. 
Thus, one finds a single-band torque formula of anisotropic superconductors as 
\cite{Kogan1988}
\begin{eqnarray}
\tau(\theta) = 
\frac{\Phi_0 {B} V}{64\pi^2\lambda^2}
\left(\frac{\gamma^2-1}{\gamma^{1/3}}\right)
\frac{\sin 2\theta}{\epsilon(\theta)} 
\ln \left( {{\gamma \eta H_{c2}^{\parallel c}} 
\over {{B}\epsilon(\theta})}\right)\,,
\label{Kogan Single Band Formula}
\end{eqnarray}
where $\epsilon(\theta)=(\sin^2\theta+\gamma^2\cos^2\theta)^{1/2}$.

Multi-band effect in anisotropic superconductors leads to the separation of the degenerated electronic anisotropy parameters into two, i.e., anisotropy in coherence length $\gamma_\xi = \xi_a/\xi_c$ and anisotropy in penetration depth $\gamma_\lambda = \lambda_c / \lambda_a$ \cite{Kogan1988,Kogan2002a}.
The magnetic field in $G$ space is assumed to be zero at $G > G_{\rm max}$ so as to fulfill the cut-off of the vortex core at $\sim \xi$ in real space.
One expresses the free energy density of Eq.~(\ref{Kogan FE}) as
\begin{eqnarray}
F =\frac{B^2}{8\pi}+\frac{\Phi_0Bm_{zz}}{32\pi^3\lambda^2m_a}\int_{G_{\rm min}}^{G_{\rm max}}\frac{dG_xdG_y}{m_{zz}G_x^2+m_cG_y^2}\,,
\label{Kogan FE-2}
\end{eqnarray}
where an upper limit of the integration $G_{\rm max}$ is fixed at 
$G_{\rm max}(\varphi)={2\pi\sqrt{\mu_am_c}}/{\xi_c\sqrt{\beta^2\cos^2\varphi+\sin^2\varphi}}\,,
$
where $\varphi$ is polar angle, $\beta^2=m_c \mu_{zz}/m_{zz}\mu_c$, $\mu_{zz}=\mu_a\sin^2\theta+\mu_c\cos^2\theta$ is a mass tensor, and  an anisotropy parameter $\gamma_\xi=\sqrt{\mu_c/\mu_a}$ as in Ref.~\cite{Kogan2002a}.
The multi-band torque formula is obtained as
\begin{eqnarray} 
\tau(\theta) = \frac{\Phi_0 {B}V}{64\pi^2\lambda^2}\frac{\gamma_\lambda^2-1}{\gamma_\lambda^{4/3}}\frac{\sin 2 \theta}{\Theta_\lambda(\theta)} \alpha(\theta)
\,,
\label{Multi-band Kogan Torque Formula}
\end{eqnarray}
where a logarithmic factor $\alpha(\theta)$ is given by
\begin{eqnarray}
\alpha(\theta)=\ln \biggl(\frac{\eta H_{c_2}^{\parallel c}}{{B}}\frac{4\Theta_\lambda(\theta)}{(\Theta_\lambda(\theta)+\Theta_\xi(\theta))^2}\biggr) \nonumber \\
-\frac{2\Theta_\lambda(\theta)}{(\Theta_\lambda(\theta)+\Theta_\xi(\theta))}\biggl(1+\frac{\mathrm{d}\Theta_\xi(\theta)/\mathrm{d}\theta}{\mathrm{d}\Theta_\lambda(\theta)/\mathrm{d}\theta}\biggr)+2 \,,
\label{Alpha in Kogan Formula}
\end{eqnarray}
$\epsilon_{\lambda}(\theta)=(\sin^2\theta+\gamma_{\lambda}^2\cos^2\theta)^{1/2}$, 
$\Theta_{\lambda}= \epsilon_{\lambda}(\theta) / \gamma_{\lambda}$, 
$\epsilon_\xi(\theta)=(\sin^2\theta+\gamma_\xi^2\cos^2\theta)^{1/2}$, and
 $\Theta_\xi= \epsilon_\xi(\theta) / \gamma_\xi$ \cite{Kogan2002a}.
Equation (\ref{Multi-band Kogan Torque Formula}) is reduced to Eq.~(\ref{Kogan Single Band Formula}) when $\gamma = \gamma_\xi = \gamma_\lambda$.
The multi-band effect is expected to take place in oxypnictides \cite{Kubota2010} and MgB$_2$ \cite{Kubota2010_1}.
As temperature decreases from $T_{c}$ down to $T=0$, a discrepancy between $\gamma_\xi$ and $\gamma_\lambda$ becomes appreciable.
Theoretically, $\gamma_\xi$ increases up to $\sim 6$ while $\gamma_\xi$ decreases down to $\sim 1$ in MgB$_2$ at $T=0$ \cite{Miranovic2003,Kogan2002b}.

It has been a long-standing issue that the ambiguity of the parameter $\eta$ is inevitable in applying Eqs.~(\ref{Kogan Single Band Formula}) and (\ref{Multi-band Kogan Torque Formula}).
The vortex-core contribution to the total energy in high-$\kappa$ superconductors is small compared to the magnetic and kinetic energy, and hence is neglected in the London model.
The energy $F$ of the vortex lattice is replaced with an integral from $G_{\rm min} \sim {2\pi}{a^{-1}}$ with an inter vortex spacing $a \sim \sqrt{{\phi_0}/{B}}$ to $G_{\rm max}\sim {2\pi}{\xi^{-1}}$ with an effective core size $\xi$.
The cutoff at $G_{\rm max} \sim {2\pi}{\xi^{-1}}$ to avoid the divergence of Eq.~(\ref{Kogan FE-2}) is an inherent shortcoming of the London approach.
The phenomenological parameter $\eta$ consists of a major factor $\eta^\prime$ and a correction factor $e^{\eta_c-1}$ as $\eta=\eta^\prime \exp (\eta_c-1)$ \cite{Kogan1996}.
The parameter $\eta^\prime$ accommodates the uncertainty in defining the core size.
The core correction $\eta_c\phi_0B/32\pi^2\lambda^2$ is added to the London free energy with an uncertain factor $\eta_c$. 
The London model is not beneficial for data analysis because it only gives not a {\it true} upper critical field $H_{\rm c2}$ but an {\it effective} upper critical field $\eta H_{\rm c2}$.
It is highly desirable to obtain the $H_{\rm c2}$ value by developing a new torque theory without containing the $\eta$ factor.

We succeeded in fixing the above-mentioned issue on $\eta$ in the preceding London models as follows.
Hao {\it et al.} \cite{Hao-Clem1991} introduced a useful cutoff function, and it was followed by Yaouanc {\it et al.} \cite{Yaouanc1997} to derive the analytical solution of the free energy density for superconductors with large Ginzburg-Landau parameter $\kappa$.
The $z$-component of the flux in the reciprocal lattice space is approximated by 
\begin{eqnarray}
h_z(\mathbf{G})\approx {\Phi_0}(1-b^4)\frac{uK_1(u)}{\Lambda_{yy}G_x^2+\Lambda_{xx}G_y^2}
\,,
\label{Yaouanc hz Eq-1}
\end{eqnarray}
where $u^2=2(\xi_x^2G_x^2+\xi_y^2G_y^2)(1+b^4)[1-2b(1-b)^2]$ and $K_1(u)$ is a modified Bessel function of the second kind.
A reduced magnetic field $b(\theta)={B}/{H_{\rm c2}(\theta)}$ as a function of $\theta$ is given by
\begin{eqnarray}
b(\theta)=({B}/{H_{\rm c2}^{\bot c}})\sqrt{\sin^2\theta+\gamma_\xi^2\cos^2\theta}\,.
\label{reduced field}
\end{eqnarray}
Extending the idea of the multiband London model of the anisotropic superconductors of Eq.~(\ref{Kogan-free-energy}) developed by Kogan \cite{Kogan2002a} we express the free energy density $F$ using the local magnetic flux $h_z(G_{pq})$ as
\begin{eqnarray}
F=\frac{B^2}{8\pi\Phi_0}\sum_{\mathbf{G}\neq 0}h_z(\mathbf{G})=\frac{B^2}{8\pi\Phi_0}\sum_{(p,q)\neq (0,0)}h_z(G_{pq})\,,
\label{Brandt FE}
\end{eqnarray}
where $G_{pq}$ is a discrete reciprocal lattice vector, $p$ and $q$ are lattice indices in a reciprocal space \cite{Kogan1981}, and $h_z(G_{pq})$ is given by
\begin{eqnarray}
h_z(G_{pq})= \frac{\sqrt{3}\Phi_0^2 (1-b^4)\epsilon_\lambda (\theta)}{2\pi^2\lambda^2\gamma_\lambda^{1/3}B(p^2-pq+q^2)}uK_1(u)\,,
\label{Yaouanc hz Eq-2}
\end{eqnarray}
using a reduced field $b(\theta)$. 
We finally obtain a novel torque formula by angular derivative as
\begin{eqnarray}
\tau(\theta)=-\frac{{B}V}{8\pi}\sum_{(p,q)\neq (0,0)}
\left[
\frac{1}{p^2-pq+q^2}\right.\cdot
\nonumber\\
\left.
\frac{\partial}{\partial\theta}\left({h_0}(1-b^4)v_{pq}K_1(v_{pq})\right)
\right]
\,,\qquad
\label{New Torque Formula}
\end{eqnarray}
where {$h_0(\theta)={\sqrt{3}\Phi_0\epsilon_\lambda}/{2\pi^2\lambda^2\gamma_\lambda^{1/3}}$}, $v_{pq}(\theta)^2=4{\pi}b(1+b^4)[1-2b(1-b)^2]$
$[\omega_{\xi\lambda}(q-p/2)^2+p^2/\omega_{\xi\lambda}]$,
and 
$\omega_{\xi\lambda}(\theta)={2\gamma_\xi\epsilon_\lambda(\theta)}/{\sqrt{3}\gamma_\lambda\epsilon_\xi(\theta)}$
are functions of $\theta$.
Equation (\ref{New Torque Formula}) is executable by numerical differentiation.
It is also possible to write down an analytical expression of Eq.~(\ref{New Torque Formula}) with the aid of the modified Bessel function of the first kind as $\partial v K_1(v)/\partial v = -v K_0(v)$.

Since our new formula of Eq.~(\ref{New Torque Formula}) does not contain a parameter $\eta$ unlike the preceding London model \cite{Kogan1988}, it is meaningful to trace the behavior of $\eta$ of the London model with respect to $\gamma$, $B$, and $H_{\rm c2}$. 
The $\eta$ is originally assumed as a factor on the order of unity, and subsequently Farrell {\it et al.} \cite{Farrell1990} clarified experimentally as $\eta\simeq 1.2 \sim 1.5$ at temperatures near $T_{\rm c}$ using the magnetization expression $M=-{\phi_0}/{32\pi^2\lambda^2}\ln {{\eta H_{\rm c2}}/{B}}$. 
The noteworthy advantage of using Eq.~(\ref{New Torque Formula}) is that one can directly obtain an upper critical field $H_{\rm c2}$ without bothering about an indefinite $\eta$ factor.
Comparison of two models yields a revisited physical interpretation of $\eta$ in connection with vortex core and $B$.

\begin{figure}[!ht]
\includegraphics[width=0.8\linewidth]{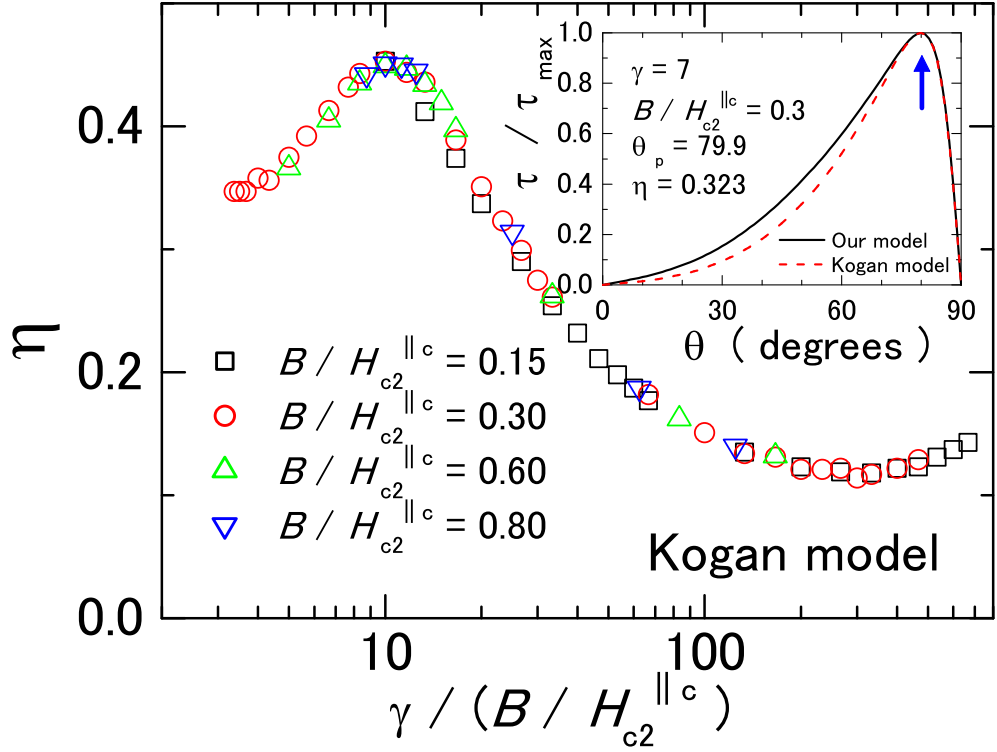}
\caption{
The parameter $\eta$ can be scaled as a function of $\gamma/({B}/H_{\rm c2}^{||c})$ on a single curve, where $\gamma$ is the anisotropy parameter, $B$ is the applied field, and $H_{\rm c2}^{||c}$ is the upper critical field parallel to the $c$-plane.
Note that $\gamma H_{\rm c2}^{||c}$ becomes the upper critical field $H_{\rm c2}^{||ab}$ parallel to the $ab$ plane.
The inset shows the torque curve of the London model (see the dashed line) of Eq.~(\ref{Multi-band Kogan Torque Formula}) for $\gamma = 7$ and {$B/H_{c2}^{\parallel c}$=0.3}, where the phenomenological parameter $\eta$ is chosen so as to give the same peak angle $\theta_p$ with our model (see solid line) of Eq.~(\ref{New Torque Formula}).
}
\label{figure1}
\end{figure}

First, we treat the single band case rather systematically where the condition $\gamma = \gamma_\lambda = \gamma_\xi$ is satisfied.
Equation~(\ref{New Torque Formula}) can be reduced to a useful analytical expression as 
\begin{eqnarray}
&&\tau(\theta)=
\frac{\sqrt{3}\Phi_0 BV}{16\pi^3\lambda^2\gamma^{1/3}}\frac{\gamma^2-1}{2\epsilon(\theta)}\sin{2\theta} \cdot
\label{Modified eq solved}
\\
&&\sum_{(p,q)\neq (0,0)}
\frac{v_{pq}}{p^2-pq+q^2}\cdot
\biggl[ (1-5b^4){K_1}(v_{pq}) 
\nonumber
\\
&& -
\frac{1-b^4}{2}\left(\frac{1+5b^4}{1+b^4}-\frac{2b(3b-1)(b-1)}{1-2b(1-b)^2}\right)v_{pq}{K_0}(v_{pq})\biggl]
\nonumber
\,.
\end{eqnarray}
We determine a peak position with respect to angle $\theta$ by using Eq.~(\ref{Modified eq solved}) when $\gamma$ and $b(\theta)$ of Eq.~(\ref{reduced field}) are given. 
We attempt to find the conditions so as to give the same peak angle $\theta_p$ by tuning a parameter $\eta$ in the Kogan model while the torque curve is normalized at the peak.
In the inset of Fig.~\ref{figure1}, the agreement of the two theories is almost perfect between $\theta_p$ and 90 degrees while it is not so good between 0 and $\theta_p$.
In the case of $\gamma=7$ and {$B/H_{\rm c2}^{||c}=0.3$}, we find $\eta=0.323$ at $\theta_p=79.9$ degrees.
We also carried out the calculations of the torque curves under the various different magnetic fields, i.e., {$B/H_{\rm c2}^{||c}$}= 0.15, 0.3, 0.6, and 0.8.
As being remarkable in Fig.~\ref{figure1}, all data in the $\eta$ versus $\gamma/({B}/H_{\rm c2}^{||c})$ representation are well collapsed into a single curve.
We consider that this is indicative of the validity of the method to compare the two theories by tuning the peak position $\theta_p$.
As $\gamma/(B/H_{\rm c2}^{||c})$ increases, $\eta$ increases at lower $\gamma/(B/H_{\rm c2}^{||c})$, it forms a peak at $\gamma/(B/H_{\rm c2}^{||c})\simeq 10$, and it decreases gradually at higher $\gamma/(B/H_{\rm c2}^{||c})$.

\begin{figure}[!ht]
\includegraphics[width=0.8\linewidth]{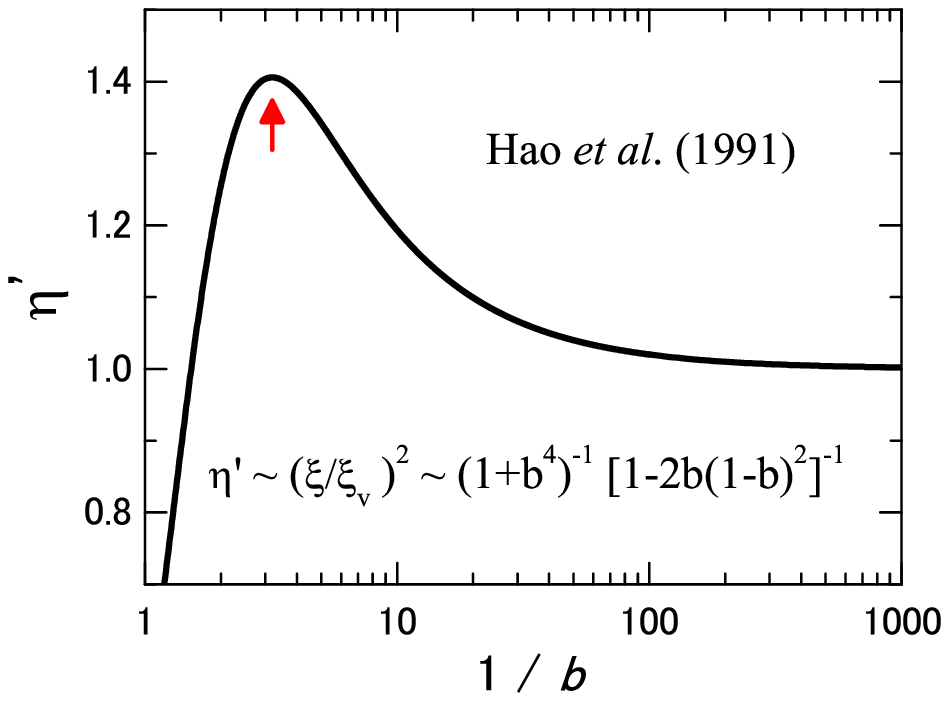}
\caption{
The qualitative behavior of $\eta^\prime$ as a function of $1/b$ expected from a theory by Hao {\it et al.} \cite{Hao-Clem1991}.
The parameter $\eta^\prime$ is a major term in the uncertainty parameter $\eta=\eta^\prime \exp (\eta_c-1)$ as given by Kogan {\it et al.} \cite{Kogan1996}, $\xi_v$ is a vortex core size, and $b$ is the reduced field.
}
\label{figure2}
\end{figure}

Second, we attempt to understand a remarkable profile of Fig.~\ref{figure1}.
Physically the inclusion of the upper critical field $H_{\rm c2}$ in the Kogan's torque formula comes from the integral interval from $2\pi a^{-1}$ to $2\pi\xi^{-1}$.
The Kogan model is based on the presupposition that the vortex core cut off always occurs at $\xi$. 
According to the theory by Hao {\it et al.} \cite{Hao-Clem1991} and by Yaouanc {\it et al.} \cite{Yaouanc1997}, however, the cut-off position is not constant at $\xi$ but varies as $\xi_v(b)$.
Therefore, the major correction parameter $\eta^\prime$ essentially comes from the replacement of $H_{\rm c2}$ by $\eta^\prime H_{\rm c2}=\eta^\prime\Phi_0/2\pi\xi^2 \approx \Phi_0/2\pi \xi_v^2=\Phi_0/2\pi \xi^2 \cdot (\xi/\xi_v)^2$.
We approximate ${\eta^{\prime}}^{-1}$ as a function of reduced field $b(\theta)$ by 
\begin{eqnarray}
\frac{1}{\eta^\prime} \sim
\left(\sqrt{2}-\frac{0.75}{\kappa}\right)^{2}(1+b^4)\left(1-2b(1-b)^2\right),
\label{Yauanc_5b}
\end{eqnarray}
where the expression of $\xi_v(b)$ was given in Refs.~\cite{Yaouanc1997,Hao-Clem1991}.
In Fig.~\ref{figure2}, we show theoretical $\eta^\prime$ as a function of $1/b$. 
The vortex core becomes that of an isolated single vortex at higher $1/b$ while it compresses at very small $1/b$ due to dense packing of vortices.
Remarkable is an enhancement of $\eta^\prime$ at an intermediate $1/b \sim 3$ (see arrow) presumably due to overlapping of vortices and resultant spreading of order parameter apart from cores.
The core correction $\eta_c$ is essentially the property of an individual vortex, and is not so dependent on $b$ and $\gamma$.
This explains qualitatively a particular behavior of $\eta$ in Fig.~\ref{figure1}.

\begin{figure}[!ht]
\begin{center}
\includegraphics[width=0.8\linewidth]{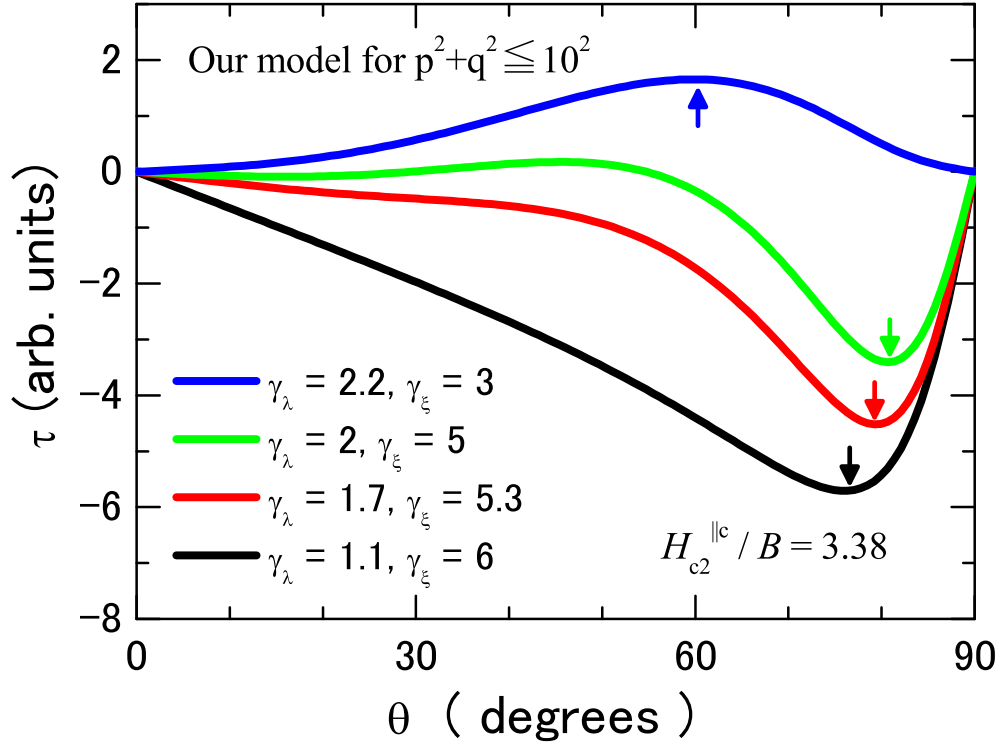}
\end{center}
\caption{
The multi-band curves are obtained by Eq.~(\ref{New Torque Formula}) for several different combinations of $\gamma_\lambda$ and $\gamma_\xi$ (see text).
The magnetic field is fixed as ${H}_{c2}^{\parallel c}/B = 3.38$ (see Ref.~\cite{Kogan2002a}).
The torque $\tau$ of Eq.~(\ref{New Torque Formula}) is given in units of $\Phi_0{B}V/64\pi^2\lambda^2$, $B$ is the applied field, $H_{\rm c2}^{||c}$ is the upper critical field parallel to the $c$-axis, and $p$ and $q$ are lattice indices in a reciprocal space.
}
\label{figure3}
\end{figure}

Third, the remarkable nature of $\eta$ of Fig.~\ref{figure1} inspires us to make a universal function $\eta(\gamma, B, H_{\rm c2})$ for substituting into the single-band Kogan model of Eq.~(\ref{Kogan Single Band Formula}). 
Consequently, the Kogan model is also able to give the upper critical field $H_{\rm c2}$ without assuming an indefinite factor of $\eta$.
Details of such an application will be published elsewhere.

\begin{figure}[!ht]
\begin{center}
\includegraphics[width=0.8\linewidth]{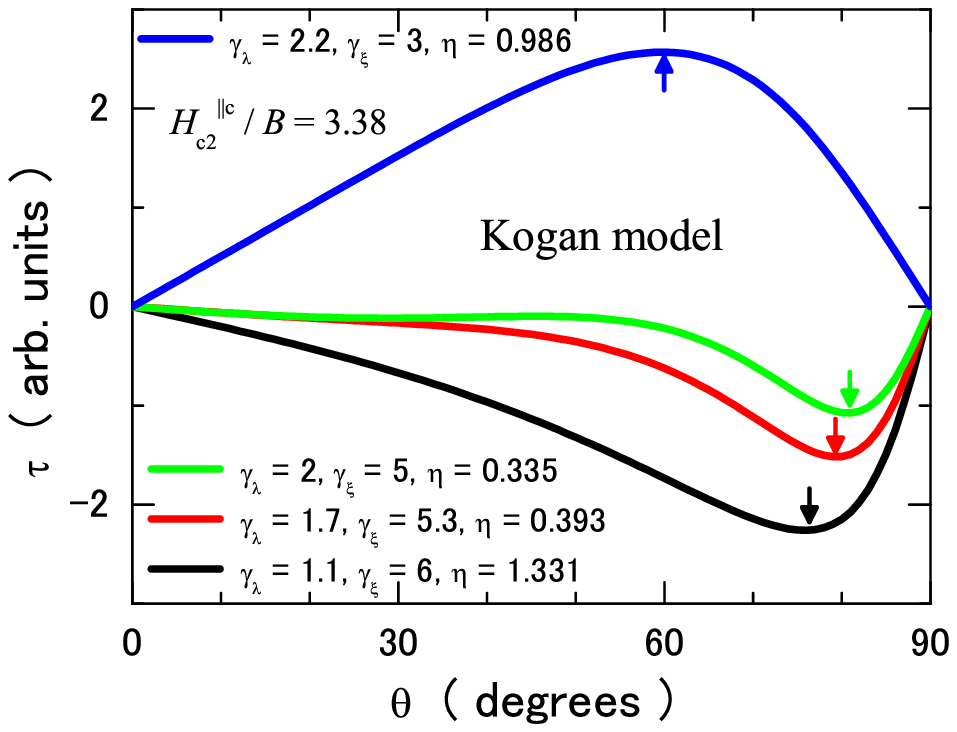}
\caption{
The torque curves using the London model Eq.~(\ref{Multi-band Kogan Torque Formula}) for several different combinations of $\gamma_\lambda$ and $\gamma_\xi$ in units of $\Phi_0BV/64\pi^2\lambda^2$  (see Figs.~1, 2, 3 of Ref.~\cite{Kogan2002a}) for comparisons).
The magnetic field is fixed as $4e^2 H_{c2}^{\parallel c}/{B} = 100$ ($e\simeq 2.71$ ), $B$ is an applied field, and $H_{\rm c2}^{||c}$ is an upper critical field parallel to the $c$-axis.
}
\label{figure4}
\end{center}
\end{figure}

Finally, we briefly discuss the generalized cases where the condition $\gamma_\xi \ne \gamma_\lambda$ is met in both our novel model as well as the Kogan model.
In Fig.~\ref{figure3}, we show a torque curve obtained by carrying out the lattice sum in the reciprocal lattice over $p^2+q^2 \le 10^2$.
We set parameters ($\gamma_\lambda = 2.2$, $\gamma_\xi = 3$), ($\gamma_\lambda = 2$, $\gamma_\xi = 5$), ($\gamma_\lambda = 1.7$, $\gamma_\xi = 5.3$), and ($\gamma_\lambda = 1.1$, $\gamma_\xi = 6$).
We reveal that summation up to $p^2+q^2 \le 10^2$ is enough to achieve a good convergence while the torque curves constructed by using the fundamental spots ($p^2+q^2\le 1^2$) is not sufficient to have a good convergence.
It is numerically confirmed that the condition $p^2+q^2 \le 10^2$ employed in Fig.~\ref{figure3} gives actually the same results as those obtained by summing up to $p^2+q^2 \le 50^2$.

The phenomenological parameter $\eta$ appeared in the Kogan model (see Eqs.~~(\ref{Kogan Single Band Formula}) and (\ref{Multi-band Kogan Torque Formula})) represents the uncertainty in defining the size of vortex core.
In Fig.~\ref{figure4}, we show the several torque curves under the different combinations of two anisotropy parameters $\gamma_\xi$, $\gamma_\lambda$, $H_{\rm c2}^{||c}/{B}$, and $\eta$. 
The parameters employed here are the same with those in Ref.~\cite{Kogan2002a}, but the condition of $\eta H_{\rm c2}^{\rm ||c}/{B}=3.38$ is replaced by $H_{\rm c2}^{\rm ||c}/{B}=3.38$.
We regard $\eta$ as a fitting parameter so as to give the same angle at the torque maximum for the case of ($\gamma_\lambda = 2.2$, $\gamma_\xi = 3$) and at the torque minimum for other cases both for Figs.~\ref{figure3} and \ref{figure4} (see arrows).
The variation of $\eta$ depending on the combination of $\gamma_\xi$ and $\gamma_\lambda$ may be related with the two competing origins in specifying $\eta$ as $\eta^\prime \exp (\eta_c-1)$ (see above).
Further systematic studies of $\eta$ as a function of $\gamma_\lambda$ and $\gamma_\xi$ are of interest as a generalized case of Fig.~\ref{figure1}.

In conclusion, we successfully elaborate the torque formula of anisotropic superconductor without containing a phenomenological parameter $\eta$.
Unlike the preceding London model \cite{Kogan1988,Kogan2002a}, we are able to estimate the {\it true} upper critical field $H_{\rm c2}$ by analyzing the torque curve.
The comparison of our new theory with the London model in the case of $\gamma_\lambda =\gamma_\xi$ has unveiled that the $\eta$ can be scaled very nicely as a function of $\gamma/({B}/H_{\rm c2}^{||c})$. 
A possible interpretation is given in view of the field dependence of the vortex core size $\xi_v$.
The behavior of $\eta$ was also investigated in the multi-band picture of $\gamma_\xi \ne \gamma_\lambda$, we found that $\eta$ changes its value but it still remains on the order of unity.

We thank M. Ichioka for suggesting the idea to explain Fig.~1 by the field dependence of the vortex core size.
We are also benefitted by discussions with M. Kato, S. Noguchi, S. Kawamata, M. Machida, T. Koyama, T. Tamegai, T. Nojima, and E. H. Brandt. 
This work was partly supported by a Grant-in-Aid for Scientific Research from the Ministry of Education, Culture, Sports, Science and Technology of Japan (Grant No. 19206104).


\begin{thebibliography}{99}



\bibitem{Yuan2009}
H. Q. Yuan, J. Singletggon, F. F. Balakirev, S. A. Baily, G. F. Chen, J. L. Luo, and N. L. Wang,
Nature {\bf 457}, 565 (2009).

%
\bibitem{Kogan1988}
V. G. Kogan, Phys. Rev. B {\bf 38}, 7049 (1988).


%
\bibitem{Kogan2002a}
V. G. Kogan, Phys. Rev. Lett. {\bf 89}, 237005 (2002).


%
\bibitem{Yaouanc1997}
A. Yaouanc, P. Dalmas de R\'eotier, and E. H. Brandt, 
Phys. Rev. B {\bf 55}, 11107  (1997).


\bibitem{Brandt1997}
E. H. Brandt, Phys. Rev. Lett. {\bf 78}, 2208 (1997).


%
\bibitem{Miller2000}
R. I. Miller, R. F. Kiefl, J.H. Brewer, J. Chakhalian, S. Dunsiger, G.D. Morris, J. E. Sonier, and W.A. MacFarlane, 
Phys. Rev. Lett. {\bf 85}, 1540 (2000).

\bibitem{Sonier1997}
J. E. Sonier, R. F. Kiefl, J. H. Brewer, J. Chakhalian, S. R. Dunsiger, W. A. MacFarlane, R. I. Miller, A. Wong, G. M. Luke, and J. W. Brill, 
Phys. Rev. Lett. {\bf 79}, 1742 (1997).

\bibitem{Sonier2005}
F. D. Callaghan, M. Laulajainen, C. V. Kaiser, and J. E. Sonier, 
Phys. Rev. Lett. {\bf 95}, 197001 (2005).

\bibitem{Kogan1996}
V. G. Kogan, A. Gurevich, J. H. Cho, D. C. Johnston, Ming Xu, J. R. Thompson, and A. Martynovich, 
Phys. Rev. B {\bf 54}, 12386 (1996).



%
\bibitem{Campbell1988}
L. J. Campbell, M. M. Doria, and V. G. Kogan, Phys. Rev. B {\bf 38}, 2439 (1988). 




\bibitem{Kubota2010}
D. Kubota, T. Ishida, M. Ishidakado, S. Shamoto, H. Eisaki, H. Kito, A. Iyo, Physica C {\bf 470}, 1109 (2010).


\bibitem{Kubota2010_1}
D. Kubota and T. Ishida, Physica C {\bf 470}, S639 (2010).



\bibitem{Miranovic2003}
P. Miranovi\'c, K. Machida, and V. G. Kogan, J. Phys. Soc. Jpn. {\bf 72}, 221-224 (2003).

\bibitem{Kogan2002b}
V. G. Kogan, Phys. Rev. B {\bf66}, 020509 (2002).

%
\bibitem{Hao-Clem1991}
Z. Hao, J. R. Clem, M. W. Mc Elfresh, L. Civale, A. P. Malozemoff, and F. Holtzberg, 
Phys. Rev. B {\bf 43}, 2844 (1991).

%
\bibitem{Kogan1981}
V. G. Kogan, Phys. Lett. A {\bf 85}, 298 (1981). 


\bibitem{Farrell1990}
D. E. Farrell, R. G. Beck, M. F. Booth, C. J. Allen, E. D. Bukowski, and D. M. Ginsberg, 
Phys. Rev. B {\bf 42}, 6758 (1990); 
D. E. Farrell, J. P. Rice, D. M. Ginsberg, J. Z. Liu,
Phys. Rev. Lett. {\bf 64}, 1573 (1990).



\end{thebibliography}
\end{document}